\documentclass[aps,prc,amsfonts]{revtex4}
\usepackage{bm,amsmath,epsfig,dcolumn,slashbox}

\begin{document}

\title{Nuclear Astrophysics}

\author{C.~R.~BRUNE}

\affiliation{Edwards Accelerator Laboratory and \\
Institute for Nuclear and Particle Physics \\
Department of Physics and Astronomy \\ 
Ohio University \\
Athens, OH 45701, USA  \\ 
brune@ohio.edu}

\begin{abstract}
Nuclear physics has a long and productive history of
application to astrophysics which continues today.
Advances in the accuracy and breadth of astrophysical data and theory
drive the need for better experimental and theoretical understanding
of the underlying nuclear physics.
This paper will review some of the scenarios where nuclear physics
plays an important role, including Big Bang Nucleosynthesis,
neutrino production by our sun, nucleosynthesis in novae,
the creation of elements heavier than iron, and neutron stars.
Big-bang nucleosynthesis is concerned with
the formation of elements with $A\le 7$ in the early Universe; the
primary nuclear physics inputs required are few-nucleon reaction
cross sections.
The nucleosynthesis of heavier elements involves
a variety of proton-, $\alpha$-, neutron-, and photon-induced reactions,
coupled with radioactive decay.
The advent of radioactive ion beam facilities
has opened an important new avenue for studying these processes,
as many involve radioactive species.
Nuclear physics also plays an important role in neutron stars:
both the nuclear equation of state and cooling
processes involving neutrino emission play a very important role.
Recent developments and also the interplay
between nuclear physics and astrophysics will be highlighted.
\end{abstract}

\maketitle

\section{Introduction}

This paper highlights some of the applications of nuclear
physics to problems in astrophysics. 
There are two primary reasons why nuclear physics plays a fundamental
role in astrophysics: (1) nuclear reactions are an important source of
energy and (2) nuclear reactions alter the isotopic composition of matter.
The close connection between these research areas has a long history.
The fields of experimental and theoretical astrophysics have
experienced considerable progress in recent years
which in turn has contributed to the vitality of nuclear astrophysics.

We now have many probes of the Universe and it's history.
The photon spectrum contains a wealth of information,
for example allowing measurements
in optical and $\gamma$-ray astronomy as well
as of the cosmic microwave background.
Neutrinos from our sun and distant supernovae have now been detected;
the prospects for future measurements from these and other sources
are most promising.
In addition important information comes from cosmic rays and
measurements of the isotopic composition of objects found
within our solar system.

Several areas of current interest in nuclear astrophysics
have recently been reviewed at the Lake Louise Winter Institute
by Hendrik Schatz\cite{Sch04} (2003)
and Richard Boyd\cite{Boy01} (2000); this review will
mostly focus on somewhat different topics.
The considerable progress in this field which has taken place
since the seminal work for Burbige, Burbige, Fowler, and Hoyle\cite{Bur57}
in 1957 is summarized in the review by Wallerstein {\em et al.}\cite{Wal97}.
An important and common theme is that the observed elemental abundances
in the universe can be understood to result from nuclear processing
in Big-Bang, stellar, and cosmic-ray scenarios.

We will begin with a discussion of Big Bang nucleosynthesis, which
is responsible for the synthesis of the lightest elements.
Then some recent nuclear physics results relating to the production
of neutrinos by our sun will be discussed. A discussion of explosive
nucleosynthesis, such as takes place in novae and X-ray bursts,
will follow next. We then cover some aspects of heavy-element
nucleosynthesis and close out with a discussion of some aspects
nuclear physics relevant to neutron stars.

\section{Big Bang Nucleosynthesis}
\label{sec:bbn}

\begin{table}[tb]
\caption{Cosmological parameters deduced from CMBR
measurements\protect\cite{Ben03}.}{
\begin{tabular}{@{}lll@{} }
\hline
$\Omega_{\rm tot}$ & (total density)          & 1.02(2) \\[1ex]
$\Omega_{\Lambda}$ & (dark energy density)    & 0.73(4) \\[1ex]
$\Omega_{\rm m}$   & (matter density)         & 0.27(4) \\[1ex]
$\Omega_{\rm b}$   & (baryon density)         & 0.044(4) \\[1ex]
$t_0$              & (age of universe)        & 13.7(2) Gyr \\[1ex]
$\eta$             & (baryon-to-photo ratio)  & $6.1(3)\times 10^{-10}$ \\[1ex]
\hline
\end{tabular}\label{tab:cmbr} }
\end{table}

Observational evidence supporting the Big-Bang model of the universe
comes primarily from four sources: (1) the Hubble expansion, (2)
the age of the universe, (3) the properties of the Cosmic
Microwave Background Radiations (CMBR), and (4) the relative
abundances of the light elements $^1{\rm H}$, $^2{\rm H}$,
$^{3,4}{\rm He}$, and $^7{\rm Li}$.
The field of Big Bang Nucleosynthesis (BBN) is concerned with the
production of these light isotopes in the early Universe.
The abundances of these
isotopes can be measured in situations where they are believed
to be primordial and compared to theoretical BBN calculations.
The main ingredients of the calculations are
the baryon density of the Universe and the rates of nuclear
reactions between the light elements. This comparison can be
used to determine the baryon density (or baryon-to-photon ratio)
of the Universe.
Nuclear physics clearly plays an important role here.
Before further delving into the
details it is worth discussing some of the exciting recent
developments in the broader field of cosmology.

\begin{figure}[b]
\centerline{\epsfxsize=3.5in\epsfbox{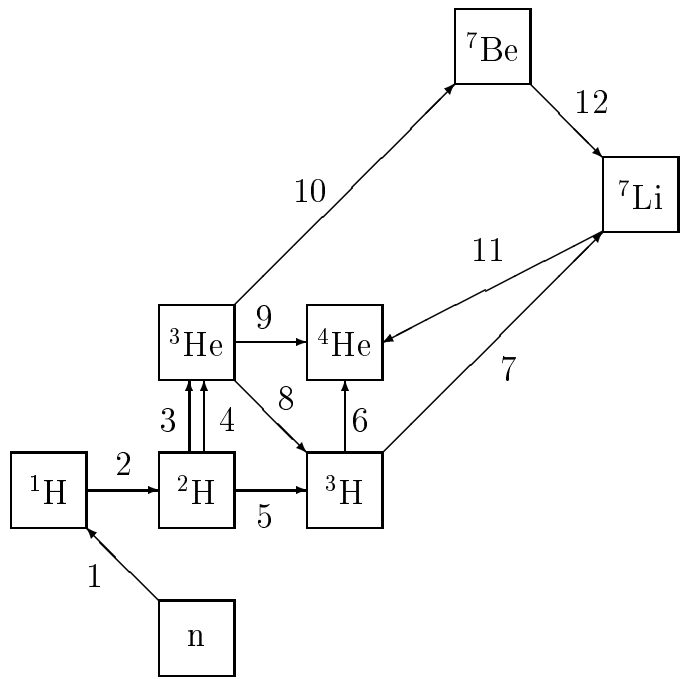}}   
\caption{The most important portion of the reaction network used
to calculate BBN abundances.
See Table~\ref{tab:bbn} for the reactions corresponding to the
numbered links.
\label{fig:bbn}}
\end{figure}

Measurements of the CMBR have reached a remarkable level of precision.
Recent results from the Wilkinson Microwave Anisotropy Probe are
reported by Bennett {\em et al.}\cite{Ben03}.
The measurements provide a wealth of cosmological information
including the age, matter density, and baryon density of the Universe;
some of their key results are summarized in Table~\ref{tab:cmbr}.
The data indicate that the baryon-to-photo ratio of
$6.1^{+0.3}_{-0.2}\times 10^{-10}$ which will be compared to the
BBN value below. Another exciting aspect of the measurements
are the findings of a non-zero cosmological constant and
non-baryonic dark matter. These findings almost certainly point to
new physics beyond standard cosmological and particle physics models.

Another source of cosmological information which has recently
become available are observations distant (high-$z$)
Type Ia supernovae\cite{Per99,Ton03}. These ``standard candles''
allow for an independent determination of the matter density
and cosmological constant. The supernova data in fact preceded the
high-precision CMBR measurements and were the first solid indication of
a non-zero cosmological constant or dark energy. This ``accelerating
expansion'' of the Universe indicates that our present understanding
of gravity is incomplete.

The physics of BBN in the present ``precision era'' has been
reviewed by Schramm and Turner\cite{DNS98}.
The nuclear reaction network is shown in Figure~\ref{fig:bbn}
and Table~\ref{tab:bbn}.
There has been considerable recent progress in the observational
determination of the primordial abundances of the light
elements. Measurements of deuterium and $^7{\rm Li}$ are
most useful for determining the baron-to-photon ratio.
Recent determinations of the the primordial D/H ratio are carried out by
measuring absorption in the direction of distant
quasi-stellar objects\cite{Ome01} and the present status of
primordial $^7{\rm Li}$/H are summarized by Ryan {\em et al.}\cite{Rya00}.

\begin{figure}[tb]
\begin{center}
\includegraphics[width=4in]{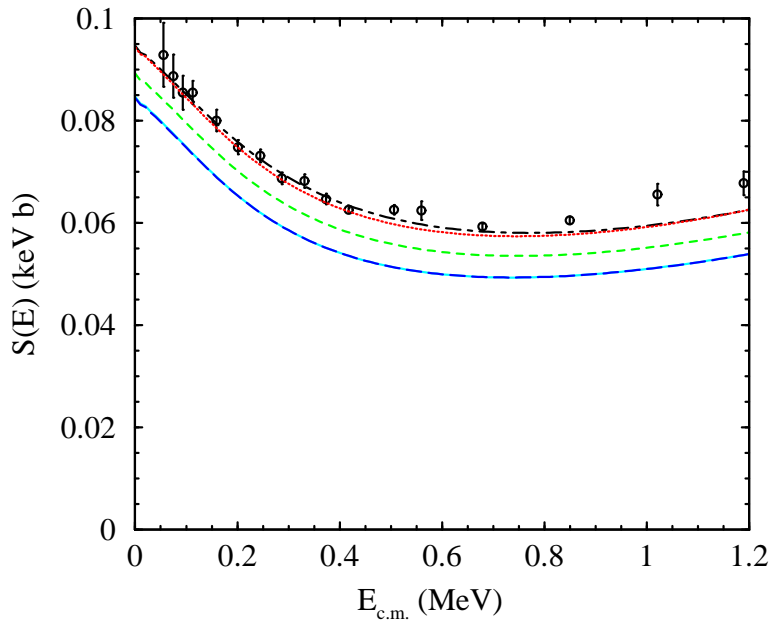}
\end{center}
\caption{The low-energy ${}^3{\rm H}(\alpha,\gamma)^7{\rm Li}$
$S$-factor. The experimental data are from Ref.\protect\cite{Bru94}
and the curves are theoretical calculations\protect\cite{Nol01}
using different assumptions to generate the scattering wavefunctions.}
\label{fig:tag}
\end{figure}

It is quite remarkable that the BBN reaction network is so simple
and requires only about a dozen nuclear reaction inputs
(in the scenarios discussed later the networks may have hundreds
of links).
The nuclear physics inputs into BBN have recently been reviewed by
Nollett and Burles\cite{Nol00}; these authors also carefully
considered how the nuclear physics uncertainties propagate in
BBN calculations. The relevant energy ranges for BBN are also given
for each reaction.
Although there has been considerable work on the
nuclear physics of the light elements over the past 50 years, these
authors point out that more work needs to be done so that nuclear
physics uncertainties do not limit cosmological conclusions.
There is also some interesting new work in the area of cross section
determinations. Recent calculations of the low-energy
$^3{\rm H}(\alpha,\gamma)^7{\rm Li}$ and
$^3{\rm He}(\alpha,\gamma)^7{\rm Be}$ cross sections have been
carried out Nollett\cite{Nol01} using realistic nucleon-nucleon
forces along with Monte Carlo techniques. These methods show
considerable promise for the future.
In Fig.~\ref{fig:tag} the calculations for the
$^3{\rm H}(\alpha,\gamma)^7{\rm Li}$ reaction are compared to
experimental data\cite{Bru94} (the $S$-factor is a
reparametrization of the cross section which approximately
removes the effect of the Coulomb barrier).
The data and calculations are seen to be in reasonable
agreement although there is some ambiguity in the calculations
depending upon how the scattering states are constructed.
Perhaps the comparison with the $^3{\rm H}(\alpha,\gamma)^7{\rm Li}$
data can be used to determine the best method for constructing
the scattering states and it can then be applied to the
$^3{\rm He}(\alpha,\gamma)^7{\rm Be}$ case where the data have
larger errors and are more discrepant.

Further work remains to be done on BBN reactions.
The $n+p\rightarrow\gamma+{\rm D}$
reaction is particularly interesting. In the energy range of
BBN ($\sim 100$~keV) there is very little data and the cross section
is in a transition region from being primarily $M1$ at low energies to
being primarily $E1$ at higher energies.
Nollett and Burles\cite{Nol00} have also pointed out that the
$^2{\rm H}(d,n)^3{\rm He}$ and $^2{\rm H}(d,p)^3{\rm H}$
reaction data needed improved accuracy.

\begin{table}[tb]
\caption{The most important reactions for BBN.
The numbers correspond to the links in Figure~\protect\ref{fig:bbn}.}{
\begin{tabular}{@{\hspace{0.5cm}}r@{\hspace{2cm}}l@{\hspace{0.5cm}}}
\hline
1 & $p \leftrightarrow n$ \\[1ex]
2 & $p(n,\gamma){}^2{\rm H}$ \\[1ex]
3 & ${}^2{\rm H}(p,\gamma){}^3{\rm He}$ \\[1ex]
4 & ${}^2{\rm H}(d,n){}^3{\rm He}$ \\[1ex]
5 & ${}^2{\rm H}(d,p){}^3{\rm H}$ \\[1ex]
6 & ${}^3{\rm H}(d,n){}^4{\rm He}$ \\[1ex]
7 & ${}^3{\rm H}(\alpha,\gamma){}^7{\rm Li}$ \\[1ex]
8 & ${}^3{\rm He}(n,p){}^3{\rm H}$ \\[1ex]
9 & ${}^3{\rm He}(d,p){}^4{\rm He}$ \\[1ex]
10 & ${}^3{\rm He}(\alpha,\gamma){}^7{\rm Li}$ \\[1ex]
11 & ${}^7{\rm Li}(p,\alpha){}^4{\rm He}$ \\[1ex]
12 & ${}^7{\rm Be}(n,p){}^7{\rm Li}$ \\[1ex]
\hline
\end{tabular}\label{tab:bbn} }
\end{table}

Comparisons between BBN calculations and observed abundances have
been performed by several recent workers\cite{Bur01,Cyb01,Cyb02}.
For the most part the conclusions are in good agreement.
Burles {\em et al.}\cite{Bur01} report a baryon-to-photon ratio of
$(5.6\pm 0.5)\times 10^{-10}$ which is in good agreement with the
value deduced from the CMBR discussed in the beginning of this
section. This concordance is considered a major triumph for cosmology.
There is however one aspect to this story which is somewhat puzzling:
the observations of primordial lithium are only marginally consistent
with the other abundances and the CMBR. If the lithium observations
are taken by themselves, they would indicated a baryon-to-photo
ratio of $\approx 3\times 10^{-10}$. Although the origin of this
discrepancy is not presently known, it may be an indication of new
physics or some as yet unknown mechanism which has universally depleted the
lithium in the old stars where it is measured.

\section{Solar Neutrinos}

Nuclear physics also plays an important role in the production and
detection of solar neutrinos. Nuclear reactions in the core of our sun
are both a source of energy and neutrinos. The relevant nuclear
physics for neutrino production has been discussed in a recent
review article\cite{Ade98}. While it is possible to analyze solar neutrino
measurements in manner which is independent of nuclear cross section
assumptions\cite{Hee96}, more information (e.g. concerning neutrino
properties) can be deduced if the cross sections are understood.
In the case of the $^1{\rm H}(p,e^+\nu_e){}^{2}{\rm H}$ reaction,
we are dependent on theoretical
calculations\cite{Bet38,Sal52,Bah68,Sch98,Par03}.
The $^3{\rm He}(^3{\rm He},2p)^4{\rm He}$,
$^3{\rm He}(^4{\rm He},\gamma)^7{\rm Be}$ and
$^7{\rm Be}(p,\gamma)^8{\rm B}$ reactions are particularly important
for calculating the flux of neutrinos from the $\beta$ decays of
$^7{\rm Be}$ and $^8{\rm B}$ which in turn supply most of the
signal measured  at Homestake (chlorine), Super Kamiokande,
and the Sudbury Neutrino Observatory.
The $^3{\rm He}(^3{\rm He},2p)^4{\rm He}$ reaction has been
measured recently in the laboratory at the relevant energies\cite{Jun98};
in the cases of $^3{\rm He}(^4{\rm He},\gamma)^7{\rm Be}$ and
$^7{\rm Be}(p,\gamma)^8{\rm B}$ the cross section measurements must be
extrapolated to lower energies. A solid theoretical and experimental
understanding of these reactions is thus required.

As discussed  by Adelburger {\em et al.}\cite{Ade98}
our present understanding of the
$^3{\rm He}(^4{\rm He},\gamma)^7{\rm Be}$ and
$^7{\rm Be}(p,\gamma)^8{\rm B}$ cross sections have not reached the
desired level of precision. It should be noted that the
$^8{\rm B}$ neutrino flux has been measured with an absolute uncertainty
of less than 4\%\cite{Fuk01}, while the predicted flux has a much
greater uncertainty ($\approx 20$\%) primarily due to
nuclear physics\cite{Bac01}. In the case of
$^3{\rm He}(^4{\rm He},\gamma)^7{\rm Be}$, further experiments are
planned, while the theoretical approach of Nollett\cite{Nol01}
discussed in Sec.~\ref{sec:bbn} helps to better understand the
extrapolation to lower energies and the relationship to the mirror reaction.
In the case of the $^7{\rm Be}(p,\gamma)^8{\rm B}$ reactions
there have been several recent direct\cite{Jun02,Bab03} and
indirect experiments\cite{Azh01,Sch03}. While these measurements
have helped the situation, there are some disagreements between
the various approaches -- the indirect approaches seem to indicate
a $\approx 10$\% lower cross section than the direct measurements.
It is hoped that additional experiments will clarify this situation.
The reader is directed to the recent papers for further discussions
and references to earlier experiments.

We will next briefly discuss the role of nuclear physics in
the detection of neutrinos via reactions on deuterium.
They Sudbury Neutrino observatory has now detected both
charged-current and neutral-current events from solar neutrinos
reacting on deuterium\cite{Ahm01,Ahm02}
which provide strong evidence for neutrino oscillations.
In order to determine the neutrino fluxes it is necessary to
to understand the $\nu+{}^2{\rm H}$ reaction cross sections.
Calculations of the charged- and neutral-current cross sections
have been carried out using two approaches.
The first ``traditional'' method\cite{Nak02} uses non-relativistic nuclear
wavefunctions generated from nucleon-nucleon potentials along with
one- and two-body current operators.
The second method approach utilizes effective field theory\cite{But01,And03}.
The two approaches agree at the level of 1\%, providing considerable
confidence in the calculated cross sections.

\section{Explosive Nucleosynthesis}

There is presently a high level of interest in understanding
the nuclear processes which take place in explosive environments
such as novae and X-ray bursts\cite{Wie98}.
In explosive situations the nuclear reactions take place very
quickly, often on a timescale much shorted than radioactive decay. 
It is then essential that nuclear reactions involving radioactive
species be taken into account. In order to address these questions
a number of radioactive-beam facilities have recently come on line
and more are planned for the future.

Novae are understood to result in binary systems in which a white
dwarf is accreting matter from its main-sequence
companion star\cite{Sta89}.
Once sufficient material has accumulated on the surface of the white
dwarf a thermonuclear runaway ensues and the accreted material
burns very quickly ($\sim10^3$~s). The luminosity of the system
increases by a factor of $\approx 10^5$ during this time.
In addition convection is predicted to bring the long-lived
$\beta^+$-emitting nuclei to the surface where they may perhaps
be seen by $\gamma$-ray telescopes.
X-ray bursters are a understood to result from a similar scenario
in which the white dwarf is replaced by neutron star.
In both scenarios the most important nuclear processes are
proton- and $\alpha$- capture reactions and $\beta^+$ decay.
The rates of these processes are needed to understand the rate of
energy production in novae and X-ray bursts as well as to
understand the elements produced. We will now focus on
novae; for more information regarding X-ray bursts the
reader is referred to the review of Schatz\cite{Sch04}.

It is thought that novae
produce a significant fraction of certain elements with $A < 30$,
e.g. $^{13}{\rm C}$, $^{15}{\rm N}$ and $^{17}{\rm O}$\cite{Jos98}.
In addition they may produce detectable fluxes of $\gamma$ rays
from the decays of $^7{\rm Be}$, $^{18}{\rm F}$, and $^{22}{\rm Ne}$.
The nuclear processing starts with the accreted material as well
material dredged up from the white dwarf. This dredge-up material
consists primarily of $^4{\rm He}$, $^{12}{\rm C}$, and
$^{16}{\rm O}$ in the case of a carbon-oxygen white dwarf and
$^{16}{\rm O}$, $^{20}{\rm Ne}$, and $^{24}{\rm Mg}$ in
the case of an oxygen-neon-magnesium dwarf\cite{Wie98}.
The reaction network generally lies along the proton-rich
side of stability. The extent of the network and the nature
of the cycling depends on the seed material and the
temperature profile.

\begin{figure}[tb]
\centerline{\epsfxsize=4.1in\epsfbox{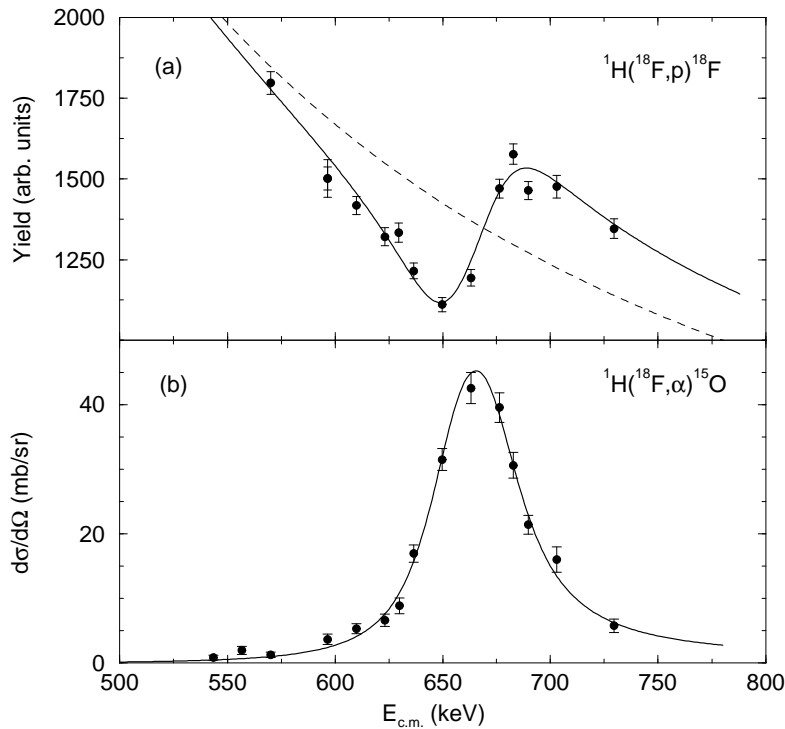}}   
\caption{The $^1{\rm H}(^{18}{\rm F},p)^{18}{\rm F}$ and
$^1{\rm H}(^{18}{\rm F},\alpha)^{18}{\rm F}$ excitation functions
measured by Bardayan {\em et al.}\protect\cite{Bar01} at ORNL-HRIBF.
The solid curves are resonance-theory fits to the data and the
dashed curve on the upper panel shows the effect of pure
Rutherford scattering.
\label{fig:hribf}}
\end{figure}

Considerable attention has recently been focused on reactions
involving the isotopes $^{17}{\rm F}$ and $^{18}{\rm F}$.
These reactions are important for determining fluxes of
511-keV $\gamma$ rays from $^{18}{\rm F}$ and
1275-keV $\gamma$ rays from $^{22}{\rm Na}$.
In addition beams of these isotopes have recently become
available at Argonne National Laboratory and the Holifield
Radioactive Ion Beam Facility at Oak Ridge National Laboratory
(ORNL-HRIBF).
The reaction $^{17}{\rm F}(p,\gamma)^{18}{\rm Ne}$ has been
studied indirectly by measuring the $^1{\rm H}(^{17}{\rm F},p)^{17}{\rm F}$
and $^{14}{\rm N}(^{17}{\rm F},^{18}{\rm Ne})$ reactions.
In the former case the excitation energy and proton width
of an important s-wave resonance were determined for the first
time\cite{Bar99,Bar00}.
While these parameters do not completely determine the
reaction rate, they are of great importance as the rate
is exponentially dependent on the resonance energy.
The $^{14}{\rm N}(^{17}{\rm F},^{18}{\rm Ne})$ proton transfer
reaction provides important information about
the structure of bound states in $^{18}{\rm Ne}$ which improves
our understanding of the (non-resonant) direct capture process\cite{Bla03}.
It now appears that the direct capture mechanism dominates the
reaction rate for nova temperatures ($<5\times 10^8$~K).
It would be highly desirable to measure directly the
$^{17}{\rm F}(p,\gamma)^{18}{\rm Ne}$ cross section, but
the cross section very small. The first step of measuring
this cross section at the peak of the aforementioned s-wave
resonance appears feasible for the ORNL-HRIBF facility.

Proton-induced reactions of $^{18}{\rm F}$ have also been studied via the
$^1{\rm H}(^{18}{\rm F},p)^{18}{\rm F}$ and
$^1{\rm H}(^{18}{\rm F},\alpha)^{18}{\rm F}$ reactions\cite{Bar01}.
As shown in Figure~\ref{fig:hribf} these measurements accurately
determine the several properties of the observed resonance:
the resonance energy $E_r=664.7\pm 1.6$~keV,
the total width $\Gamma=39.0\pm 1.6$~keV,
the proton width $\Gamma_p/\Gamma=0.39\pm 0.2$,
and the $(p,\alpha)$ resonance strength $\omega\gamma=6.2\pm 0.3$~keV.
The reaction rate for the $^{18}{\rm F}(p,\alpha)^{15}{\rm O}$
reaction is now known within 10\% for $0.4 \le T \le 2.0$~GK,
but is still uncertain for lower temperatures due to the
uncertainties associated with low-energy resonances.
Although $\Gamma_\gamma$ for the 665-keV resonance has net yet been
determined it can be concluded that the
$^{18}{\rm F}(p,\alpha)^{15}{\rm O}$ reaction is much faster than
the $^{18}{\rm F}(p,\gamma)^{19}{\rm Ne}$ in both novae and X-ray bursters.

The best opportunity for observing a nuclear $\gamma$-ray line
produced by a nova appears to be with the 1275-keV line from $^{22}{\rm Na}$.
The $\gamma$-ray flux prediction depends on several reaction rates,
including the proton-induced reactions of fluorine isotopes mentioned
above. In addition, the $^{21}{\rm Na}(p,\gamma)^{22}{\rm Mg}$
reaction plays a very important role (note that $^{22}{\rm Mg}$
decays into $^{22}{\rm Na}$). Model calculations of
oxygen-neon-magnesium novae\cite{Jos98} indicate the
uncertainty in the $^{21}{\rm Na}(p,\gamma)^{22}{\rm Mg}$
reaction rate is the dominant source of uncertainty in
calculated 1275-keV $\gamma$-ray flux. Fortunately the strength
of the key 206-keV resonance in this reaction was recently
measured at the TRIUMF-ISAC facilty\cite{Bis03}.
These measurements support the prediction that maximum detectability
distance for the European Space Agency's INTEGRAL spectrometer (SPI)
is approximately 1~kpc\cite{Gom98}.

\section{Nucleosynthesis of Elements Heavier than Iron}

The more abundant heavy nuclei are primarily synthesized by neutron capture
reactions. These processes can be categorized into the ``slow''
and ``rapid'' neutron capture process, depending on how the
neutron-capture timescale compares to the $\beta^-$ decay time.
These mechanisms are dubbed the s-process and r-process, respectively.
Some proton-rich isotopes are thought to be synthesized by a succession of
rapid proton captures (rp-process). In addition other mechanisms
including photodissociation and neutrino
spallation also play a role\cite{Boy01}.
Although certain parameters can be deduced from the observed
abundances, the sites of these nucleosynthetic processes is not
always clear. Many of these scenarios are likely associated with
core-collapse supernovae; however a unified description
of these phenomena is still a work in progress.

\subsection{S-Process}

The main component of the s-process ($A>90)$
is relatively well-understood to occur in
low-to-intermediate-mass [$(1.5-8)M_\odot$]
thermally-pulsing asymptotic giant branch (AGB) stars\cite{Gal03}.
The so-called weak component is produced in
heavier $\sim 25M_\odot$ stars which are burning helium in their cores.
At the present time research in this area is focused on two
major areas: (1) measurements of neutron capture cross sections and
(2) better determinations of the neutron source reactions.

Particular attention is being focused on neutron capture cross
sections on {\em unstable} isotopes in the s-process path.
In some of these cases neutron capture can compete with $\beta^-$ decay.
These branch points can provide important information about the
neutron density during the s-process. The measurements are difficult
because radioactive targets are involved, but efficient techniques for
measuring cross sections with small samples are under
development\cite{Kap03}.

The neutrons for the s-process are produced via the
$^{13}{\rm C}(\alpha,n)^{16}{\rm O}$ and
$^{22}{\rm Ne}(\alpha,n)^{25}{\rm Mg}$ reactions.
At the temperatures involved in the s-process
[$T\approx (1-3)\times 10^8$~K] the corresponding energies
of the colliding nuclei are far below the Coulomb barrier
and the cross sections are very small very difficult
if not impossible to measure directly. More experimental and
theoretical work is needed to determine the rates
of these reactions at astrophysical energies\cite{Hal97,Hei01,Jae01}.

\subsection{R-Process}

The r-process is responsible for the peaks in the solar-system
abundances seen at $A\approx 130$ and~160. These peaks are thought
to result from the neutron shell closures and $N=82$ and~126.
Nuclei consisting of a closed neutron shell plus one neutron
have very low neutron separation energies and are much more
likely to undergo $(\gamma,n)$ reactions than further neutron captures.
Considerations of statistical equilibrium thus favor the build-up
of abundances at the closed neutron shells. As the high-temperature
environment cools the reactions fall out of equilibrium and the
nuclei $\beta^-$ decay back to the stability line. Although
this general framework has been understood for a long time, the
details concerning nuclear physics and the astrophysical
site remain largely unknown.

Several possible sites for the r-process have been put forward,
including neutron star mergers\cite{Ros99}, the ejecta of core-collapse
supernovae heated by neutrino wind\cite{Wos92},
and in the accretion disks of neutron stars following core-collapse
supernovae\cite{Cam01}. At the present time the site remains an
open question. Measurements of the abundances of several
r-process elements have recently
been carried out in ultra-metal-poor halo stars\cite{Sne00}.
These stars show an r-process abundance pattern which closely matches the
(scaled) solar r-process abundances, indicating that
the r-process may be universal in nature.

Hardly any of the isotopes in the projected path of the r-process
have been studied in the laboratory. Knowledge of the neutron separation
energies and half-lives are critical for determining the nature
of the statistical equilibrium. The emission of $\beta$-delayed
neutrons, fission, and the neutron-capture cross sections will impact the
final abundance pattern\cite{Pan03}.
Depending upon the site, neutrino oscillations may also have
a very significant effect on the r-process\cite{Ful01}.
A new facility is has been proposed to be built in the United States,
the Rare Isotope Accelerator, would be able to produce the
majority of the isotopes expected to be involved in the r-process.
In addition to measurements of $Q$-values and decay properties,
other nuclear structure information such as level densities
and $\gamma$-ray strength functions can be determined which are
important for estimating neutron capture cross sections\cite{Gor03}.

\section{Neutron Stars}

Nuclear physics also plays an important role in neutron star interiors.
Of particular interest
are the equation of state and cooling by neutrino emission.
The nuclear equation of state largely determines the mass versus radius curve
for neutron stars\cite{Lat01}. 
Since the core density of neutron stars is several times normal
nuclear density, this environment provides unique tests of
nuclear physics which cannot be carried out in terrestrial laboratories.
In principle the mass and radius are
measurable so that the equation of state can be tested. At present
many neutron star masses are accurately known, but there are no
radius determinations -- although there is promise for future
measurements. Interestingly the vast majority of measured neutron
star masses are clustered around 1.4$M_\odot$ which may be
an indication that neutron stars are usually formed with near-maximal
mass. Theoretical calculations\cite{Gle01} indicate that the
maximum possible neutron star mass is in the range $(1.5-1.7)M_\odot$
even if rather exotic nuclear physics is allowed for.
In light of the fact that radius determinations will soon be
possible, the prospects for an observational constraint on
the nuclear equation of state are good\cite{Pra03}.

The ages and temperatures of neutron stars can be determined in
many case by the morphology of the supernova remnant
and x-ray spectra, respectively.
The age versus temperature trajectory is closely related to
cooling rate of neutron stars.
Of particular interest here are the neutrino-producing processes\cite{Yak01}
\begin{eqnarray}
n+n &\rightarrow& n+p+e^-+\bar{\nu}_e \nonumber \\
n+p &\rightarrow& p+p+e^-+\bar{\nu}_e, \nonumber \\
n+n &\rightarrow& n+n+\nu_x+\bar{\nu}_x, {\rm ~and} \nonumber \\
n+p &\rightarrow& n+p+\nu_x+\bar{\nu}_x . \nonumber
\end{eqnarray}
Many-body effects are very important due to the extremely high
densities in neutron star cores. It is very likely that the
neutrons are in a superfluid state\cite{Sca98}, and the effects
of strong magnetic fields may be important\cite{Sca98,van00}.
It is expected that the fruitful interplay between astronomy
and few-nucleon dynamics will continue in the future
as better observations and calculations become available.

\section{Conclusions}

Some applications of nuclear physics to Big Bang Nucleosynthesis,
solar neutrinos, explosive nucleosynthesis, the production of
elements heavier than iron, and neutron stars have been reviewed.
The future in all of these areas is very exciting as
improved measurements of the elemental abundances, solar
neutrino fluxes, galactic $\gamma$ rays,
and neutron star properties are planned.
On the nuclear physics side new facilities such as the
Rare Isotope Accelerator
will improve our understanding of the underlying nuclear physics.
Together these efforts will lead to a greater understanding
of both our Universe {\em and} nuclear physics.

\section{Acknowledgments}

I would like to thank the organizers of this conference for the
opportunity to attend the Lake Louise Winter Institute and
present these lectures.
I also thank K.M.~Nollett for supplying Fig.~\ref{fig:tag}.
This work was supported in part by the U.S. Department of 
Energy, under Grant No. DE-FG02-88ER40387.

\end{document}